\documentclass[aps,prd,preprintnumbers,showpacs]{revtex4}
\usepackage[dvips]{graphicx}
\begin{document}

%
%

\eprint{Nisho-2-2018}
\title{Explanation of Detailed Spectral Properties of FRBs by Axion Star Model }
\author{Aiichi Iwazaki}
\affiliation{Nishogakusha University,\\ 
6-16 Sanbantyo Chiyoda Tokyo 102-8336, Japan.}   
\date{Oct. 10, 2018, last revised Jul. 28, 2019}

\begin{abstract}%
We have proposed a generation mechanism of non repeating ( repeating ) fast radio bursts: They arise by
axion star collisions with neutron stars ( accretion disks of galactic black holes ).  
The axion star as coherent state of axions with mass $m_a$ generates homogeneous electric
field oscillating with frequency $m_a/2\pi$ under strong magnetic fields.
The field makes electrons coherently oscillate
and emit the coherent dipole radiations ( FRBs ). The radiations stop
when the oscillations are disturbed by thermal fluctuations produced by the themalization of the oscillating energies.
Thus, the durations of the FRBs are determined by the time scale of the thermalization.
We show that it is much shorter than $1$ms.
Line spectra of the dipole radiations are broadened by the thermal effects.
The thermally broaden spectra have a feature that the bandwidths $\delta\nu$ are proportional to their center frequencies $\nu_c$;
$\delta\nu \propto \nu_c$. 
Because the accretion disks can orbit with relativistic velocities, the radiations are
Doppler shifted. It leads to the presences of various center frequencies ( $1.2$GHz$\sim $$7$GHz ) in repeating FRB 121102.
On the other hand, non repeating FRBs do not show such variety of the center frequencies. They come from the surfaces of neutron stars
whose motions are non relativistic. 
The Doppler shift also makes the durations of the bursts with higher frequencies
become shorter. Because the magnetic fields of the neutron stars are supposed to be stronger than those of the accretion disks,
the peak flux densities of non repeating FRBs are larger than those of repeating FRB 121102. The strong magnetic fields
of the neutron stars also lead to much wide bandwidths of non repeating FRBs, which are over the extent of the receiver frequency range.
The spectral features of the recently discovered new repeating FRB 180814.J0422+75 are coincident with our
general analyses of the repeating FRB 121102. 
\end{abstract}

\hspace*{0.3cm}

\hspace*{1cm}


\maketitle
\section{Introduction}

Since a Fast Radio Burst ( FRB) has been originally reported\cite{frb},
more than $50$ FRBs have been observed.
Among them, only a FRB 121102 emit bursts repeatedly\cite{rep,rep1,repmulti,repmulti1} and so is named as repeating FRB.
( CHIME has very recently observed\cite{chime} a new repeating FRB 180814.J0422+75. )
The detailed follow up observations of the FRB 121102 have been performed. As a consequence,
we have obtained detailed spectral and temporal properties of the FRB 121102 in a wide frequency range $1$GHz$\sim$$8$GHz.
On the other hand,
the other FRBs have not been observed to repeat and so are named as non repeating FRBs.
Thus, their observations have been limited only in a frequency range $0.8$GHz$\sim$$1.6$GHz.
Although observations \cite{low} at low frequencies $145$MHz, $182$MHz and $350$MHz have been performed, any FRBs have not been detected.

The typical properties common in both types of the FRBs are in the following.
The durations are 
a few milliseconds. They have the large dispersion measures
suggesting 
extra-galactic origins  so that 
the large amount of the energies $\sim 10^{40}$erg/s are
produced at the radio frequencies. Additionally,
the event rate of non repeating FRBs is approximately 
$\sim 10^3$ per day at the earth.

These fundamental properties can be explained by  
many models\cite{katz,frbmodel} for sources of FRBs; magnetar, neutron stars merger, highly rotating newly born neutron stars, etc.. 
To restrict the models,
we need to explain spectral and temporal properties of FRBs in detail. Especially, the detailed observations of
the repeating FRB 121102 help to select a few models among them.

\vspace{0.1cm}
The follow up observations\cite{re3G,re1.6G,op} have shown that the FRB 121102 is originated 
in a dwarf galaxy with redshift $z\simeq 0.19$. The galaxy is conjectured to be an AGN.
The bursts of the FRB 121102 have been observed\cite{repmulti,repmulti1,4-8,4-5} 
with multiple bands $1.2$GHz $\sim 1.5$GHz, $1.7$GHz $\sim 2.3$GHz, $2.5$GHz $\sim 3.5$GHz, $4.5$GHz$\sim 5$GHz and
$4$GHz$\sim 8$GHz.
Consequently, we have found that the bursts exhibit finite bandwidths; they are not broadband.
Sometimes it is stated that the spectra have structures. But we would like to state that the bursts are narrowband.
This is because simultaneous multi band observations clearly show that any repeating bursts have no spectra at lower frequencies $<100$MHz
and at much higher frequencies $>13$GHz. ( Our axion model\cite{iwazaki,iwazaki2} for FRBs predicts that any FRBs exhibit thermally broaden narrow band spectra
although they intrinsically exhibit an line spectrum. ) 
In addition, the bursts have no visible light, X ray or gamma ray components\cite{light,x}, which have been confirmed by
simultaneous observations. 
Although there is a possibility of non detections owing to low sensitivities of telescopes for the electromagnetic waves at such low or high frequencies,
the observations of the repeating FRB 121102 indicate that the bursts are really narrowband just as 
molecular line emissions.
Our axion model predicts such narrow bandwidths of the non repeating and repeating bursts, although
the bandwidths of the non repeating FRBs are much wide to be over the extent of receiver frequency ranges.

Along with the narrow bands of the repeating FRB,
the observations exhibit a tendency that the burst widths $\delta\nu$ are proportional to the center frequencies $\nu_c$; $\delta\nu \propto\nu_c$.
For instance, we can see bandwidths roughly given by $\sim 500$MHz at the center frequencies $\sim 3$GHz\cite{repmulti1} and similarly
$\sim 1$GHz at the frequencies $\sim 6$GHz\cite{4-8}.
It seems that the proportionality holds even for the bursts with the frequencies $\sim 2$GHz\cite{repmulti} ( their bandwidths $\sim 300$MHz ) and 
$1.3$GHz\cite{rep1} ( bandwidths $\sim 200$MHz ). The recent observation\cite{chime} by the CHIME shows that the repeating FRB 180814.J0422+75
has bandwidth $\sim 100$MHz at the center frequency $\sim 700$MHz.

Furthermore, it is recognized that the peak flux densities of non repeating FRBs are approximately ten times higher than those of repeating FRBs.
More interestingly it has recently been pointed out\cite{4-8} that in the FRB 121102
the durations of the bursts with higher frequencies are shorter than the ones of the bursts with lower frequencies.
It seems that the product of the duration and the center frequency is approximately constant; it is independent of
the frequencies $\nu_c$. The feature also holds even for the new repeating FRB 180814.J0422+75.

\vspace{0.1cm}
The follow up observations clarify the presence of a persistent radio source spatially coincident with the bursting source of the FRB 121102. 
The persistent source might be nebula associated with neutron star or AGN. The presence of the persistent source
has been considered as a main evidence supporting newly born magnetor model for the FRB 121102.

Rotation measures of FRBs have also been obtained\cite{rotation,lrotation}. 
The radiations of the FRB 121102 are $100$\% linearly polarized. 
It is notable that the rotation measures\cite{lrotation} of the FRB 121102 are very large such as $\sim 10^5$rad m$^{-2}$, while
those of the non repeating FRBs are at most $\sim 2\times10^2$rad m$^{-2}$. It indicates that the environment
of the FRB121102 is similar to the one of PRS J1745-2900 located near the black hole 
of the Milky Way galaxy, 
while those of the non repeating FRBs are similar to environments of ordinary pulsars ( neutron stars ) in the Milky Way galaxy.
The observations of the large rotation measure and the persistent source in the vicinity of the source of the FRB 121102
lead to a conjecture that the source of the FRB 121102 might be a newly born magnetar or would be associated with AGN. 

\vspace{0.2cm}
These spectral and temporal properties of the repeating FRB 121102 are clue to find sources of the fast radio bursts.
In the paper we explain all of the properties by using the axion model\cite{iwazaki,iwazaki2}; axion star collides with neutron stars ( source of non repeating FRBs ) 
and with accretion disks
of galactic black holes ( source of repeating FRBs like FRB 121102 ).
In the axion model, FRBs are emitted from electron gases in these astrophysical objects which the axion star collides.
In particular, accretion disks
can orbit with relativistic velocities when they are in the vicinity of the black hole.
Thus, we can expect some phenomena associated with Doppler effects. We show that
the spectral and temporal properties mentioned above are caused by the Doppler effects.
On the other hand, the electron gases of neutron stars
move only with non relativistic velocities. The fact leads to non observations of the non repeating FRBs with low frequencies ( $<200$MHz ) or high frequencies ( $>3$GHz ).
Probably, magnetic fields of neutron stars are stronger than those of accretion disks. Owing to the fact, 
peak flux densities of the non repeating FRBs are larger than
than those of the repeating FRB. 
Additionally, it may cause too large bandwidths ( $>1$GHz ) of non repeating FRBs to be detectable; they
are over the extent of the receiver frequency ranges. Indeed,
their bandwidths have not been detected. ( Very recent observation\cite{finite} exhibits that spectra of non repeating FRBs have no components with low frequencies. 
It suggests that the bursts are not broadband like the repeating FRB 121102 and their 
bandwidths are larger than those of the repeating FRBs with frequencies $\sim 1.4$GHz. ) 
These detailed spectral and temporal properties mentioned above are not clearly explained by the newly born magnetar ( neutron star ) model. 

\vspace{0.1cm}
( We assume in the paper that the accretion disk is geometrically sufficiently thin in order for the axion star to be able to pass the disk and
to collide it several times orbiting a black hole.   
Such a disk would have much low temperature and have 
strong magnetic fields $\sim 10^{10}$G.
Actually, there is a model\cite{accretion,review} of geometrically thin accretion disk with the strong magnetic fields 
in the vicinity of the black hole.  Furthermore, we assume that 
the dark matter in the Universe is mainly composed of axion stars, which 
highly condense in the centers of galaxies and frequently collide the accretion disks. )

\vspace{0.1cm}

The axion model predicts coherent emissions at radio frequencies. Additionally, the model predicts
some ejecta in the axion star collision with neutron stars or accretion disks.
In particular, these ejecta repeatedly take place in the repeating FRB 121102. The ejecta would form the persistent radio source or could power
the source observed in the vicinity of the FRB 121102.

\vspace{0.1cm}
In the next section(\ref{2}), we show the production mechanism of the coherent radiations. 
The axion stars are coherent states of axions. We present approximate solutions of gravitationally loosely bound states of the axions in the section(\ref{3}).
Although they are spherical, actual forms of the axion stars are extremely distorted\cite{iwazaki4} by tidal forces. But the coherence of the axions is kept
because the number of the axions in the volume $m_a^{-3}$ is extremely large such as $\sim 10^{40}$.
Our analyses hold even for such non spherical axion stars\cite{iwazaki2}. This is because the tidal forces do not change the masses of the axion stars
and the parameters ( especially, the amplitudes of the axion fields )  
characterizing the axion stars are almost fixed by their masses. 
These parameters are used for the evaluation of
the emission properties of FRBs. Therefore, our spherical solutions themselves presented in section(\ref{3}) are not essential for
the explanation of various properties of FRBs. In this section, we discuss how
the tidally deformed axion star collide accretion disks, noting that their motions are described by the Kepler motion.
In the section(\ref{4})
we show that a sufficiently large amount of radiation energies is produced 
from the axion star collisions with neutron stars or accretion disks of galactic black holes so as to be coincident with the observations. 
We point out spectral properties of repeating FRBs and explain these features with our axion star model
in the section(\ref{5}). In the final section(\ref{6}) we summarize our results.

\section{Coherent Dipole Radiations}
\label{2}
FRBs are coherent radiations because their energy fluxes are extremely large $\sim 10^{40}$erg/s and 
the sizes of their sources are small $\sim 1$ms$\times 10^{10}\rm cm/s=10^7$cm. 
It is expected that they are emitted in environments with strong magnetic fields such as neutron stars. 
But, it is difficult to make explicit the emission mechanism of coherent radiations associated with the strong magnetic fields.
It is believed\cite{frbmodel} that some non linear effects in the plasma with strong magnetic fields generate the coherent radiations.

\vspace{0.1cm}
We show that when axion stars collide with magnetized electron gases, the electrons emit coherent dipole radiations.
The mechanism\cite{iwazaki2} is very simple.
First, we notice that electrons can coherently emit radiations when they are imposed by spatially homogeneous oscillating electric 
field $\vec{E}(t)=\vec{E}_0\cos(m_a t)$.
( Such an electric field is generated by axion stars when they are imposed by strong magnetic fields, as we show in next section. )
The motion of each electron with electric charge $e$ is described by the momentum $\vec{p}(t)=e\vec{E}_0\sin(m_a t)/m_a+\vec{p}(t=0)$.
Obviously, electrons coherently oscillate and emit dipole radiations with the frequency $m_a/2\pi$. 
( Electrons in a region where the homogeneous electric field is present can emit coherent radiations. 
Thus, the number $N_e$ of electrons can coherently emit the radiations so that their flux 
is large proportional to $N_e^2$. It is naively expected that the electrons in the volume such as $(2\pi/m_a)^3$
coherently emit the radiations, where $2\pi/m_a$ denotes the Compton wave length of the axion. )
The radiations are monochromatic and are $100$\% linearly polarized.
The most of the radiations are emitted in the direction perpendicular to the electric field.
This is the emission mechanism of the coherent radiations in the axion model.
( The conversion of the axions into radio waves in the magnetosphere of neutron star has been discussed\cite{conversion}. 
As a result, the radio waves with the frequency $\nu_{\rm int}$ are produced.
The conversion resonantly arises when the plasma frequency is equal to the frequency $\nu_{\rm int}$. But the radiations are incoherent so that
their energies are extremely smaller than the energies of the FRBs or the whole energy of the axion star. 
Thus, the axion star survives against the conversion even if 
it passes the regions with the plasma frequencies larger than $\nu_{\rm int}$. )

\vspace{0.1cm} 
In general,
the coherent oscillations are disturbed by thermal fluctuations. The momenta of the electrons at temperature $T$ are given by $\vec{p}(t)+\vec{p}_{\rm thermal}$
with $\vec{p}_{\rm thermal}^2/2m_e=T$ as long as $p(t)\gg p_{\rm thermal}$; $m_e$ denotes the electron mass.
Thus, the coherent oscillations are not disturbed by the thermal fluctuations at low temperatures. But,
the temperature of the electron gases increases because the oscillation energies are thermalized.
Once the temperature goes beyond a critical one $T_c$,
the coherent oscillations are disturbed so that the coherent radiations stop. Such a critical temperature is approximately 
given by $T_c\simeq \vec{p}(t)^2/2m_e\simeq (eE_0/m_a)^2/2m_e$.
The temperature $T_c$ determines the widths of the thermally broadened spectra of the dipole radiations.
We show later that the time scale needed for the thermalization is much less than the observed durations ( millisecond ) of FRBs. 

\section{Axion Stars and Generation of Electric Field}
\label{3}
\subsection{Axions and Axion Stars}
Before showing how the electric fields are generated by axion stars, we explain what is axion or axion star. 
The axion is the Nambu-Goldstone boson\cite{axion} associated with U(1) Pecci-Quinn symmetry. 
The symmetry is chiral and naturally solves the strong CP problem in QCD: The problem is why the CP violating term $G_{\mu,\nu}\tilde{G}^{\mu,\nu}$ is
absent in QCD Lagrangian where $G_{\mu,\nu}$ ( $\tilde{G}^{\mu,\nu}$ ) denotes ( dual ) fields strength of gluons.
( If the exact chiral symmetry is present, the term can be made to vanish. But it is no so because u and d quarks have small bare masses. )
Although the axion is described by a real massless scalar field, it acquires 
its mass $m_a$ through chiral anomaly because the Pecci-Quinn symmetry is chiral. 
Thus, instantons in QCD give rise to the mass of the axion. 

The axions are one of most promising candidates of
the dark matter. 
The axions may form axion stars
known as oscillaton\cite{axion2,osci}; the axion stars are gravitationally bounded states of axions.
In the early Universe,
axion miniclusters\cite{kolb}
are produced after the QCD phase transition and 
become the dominant component of the dark matter.
After the formation of the axion miniclusters, the axions condense
and form axion stars by gravitational cooling\cite{cooling,osci} as more compact coherent states of axions.
In the present paper we assume that the dark matter is mainly composed of the axion stars.

\vspace{0.1cm}
As the axion field is real scalar, 
there are no static solutions in a system of the axion coupled with the Einstein gravity.
Only oscillating solutions are present, which represent
oscillating coherent states of the axions.
When the mass $M_a$ of the axion star is small enough
for the binding energies of the axions to much less than the axion mass $m_a$,  
approximate spherical solutions\cite{iwazaki3} are given by 

\begin{equation}
\label{a}
a=a_0f_a\exp(-\frac{r}{R_a})\cos(m_a t) \quad \mbox{with} 
\quad a_0=0.8\times 10^{-7}\Bigl(\frac{M_a}{2\times 10^{-12}M_{\odot}}\Bigr)^2\Big(\frac{m_a}{0.6\times10^{-5}\mbox{eV}}\Big)^3
\end{equation}
with the decay constant $f_a$ of the axions, 
where the radius $R_a$ of the axion stars is approximately given by
\begin{equation}
R_a=\frac{1}{GM_a m_a^2}\simeq 
360\mbox{km}\Bigr(\frac{0.6\times10^{-5}\mbox{eV}}{m_a}\Bigl)^2\frac{2\times10^{-12}M_{\odot}}{M_a}
\end{equation}
where $G$ denotes the gravitational constant.
The decay constant $f_a$ is related with the mass $m_a$; 
$m_a\simeq 6\times 10^{-6}\mbox{eV}\times (10^{12}\mbox{GeV}/f_a)$. There are two unknown parameters $m_a$ and $M_a$ ( or $R_a$ ). 
The mass $M_a$ ( or radius $R_a$ ) of the axion star was obtained\cite{iwazaki,rate} by comparing the collision rate between neutron stars
and axion stars in a galaxy with the event rate of FRBs $\sim 10^{-3}$ per year in a galaxy.
It was found that the mass takes a value roughly given by $M_a=(10^{-11}\sim 10^{-12})M_{\odot}$.
The axion mass $m_a$ is supposed to be $\simeq 0.6\times10^{-5}$eV 
so as for the intrinsic frequency $\nu_{\rm int}$ of the non repeating FRBs to be given by
$m_a/2\pi\simeq 1.4$GHz. The value is in a window allowed in axion searches. 
( The observed frequencies $\nu$ receive effects of cosmological expansions such that $\nu=\nu_{\rm int}/(1+z)$ with redshift $z$. ) 
The spherical solutions represent gravitationally loosely bound states of the axions.
They are easily deformed by tidal forces of neutron stars or black holes.
The solutions themselves are not used for the estimation of electric fields generated by the deformed axion stars. 
But the value $a_0$ can be
approximately used for the estimation in the following reason,
even if the spherical forms of the solutions are deformed.

\vspace{0.1cm}

The above solutions represent
coherent states of the axions possessing much small momentum $k$ given by $k\sim 1/R_a\ll m_a$.
The number of the axions in the volume with $m_a^{-3}$ ( $m_a^{-1}$ is Compton wavelength of the axion ) is extremely large; $M_a/m_a\times(1/R_am_a)^3\sim 10^{41}$.
It turns out that the coherence is very rigid. That is, the coherence is kept even if
tidal forces of neutron stars or black holes distort the shapes of the axion stars like long sticks\cite{iwazaki4}. 
It implies that the deformed axion stars can be treated classically just as classical electromagnetic waves.  
Because the tidal forces do not change the mass $M_a$ of the axion stars, 
we may approximately estimate the value of $a_0=a/f_a$ when the axion stars are deformed, by using the formula 
$M_a=\int d^3x ((\partial_t a)^2+(\vec{\partial}_x a)^2+(m_aa)^2)/2\sim (m_a a)^2\times \mbox{``volume of the axion star''}$. 
The tidal forces extremely deform the spherical forms and
also change the volumes of the axion stars. But even if the volume becomes $10$ times larger, the value $a$ only changes to be
$3$ times smaller. Therefore, we may approximately use the value $a_0$ in eq(\ref{a}) of the spherical solutions, 
in order to estimate the strength of the electric field $\vec{E}$ generated by the axion star collisions
with neutron stars or accretion disks.

\vspace{0.1cm}
\subsection{Generation of Electric field under external Magnetic Field} 
Now we show how the axion star generates electric field under magnetic fields.
It is well known that
the axion $a(\vec{x},t)$ couples with both electric $\vec{E}$ and magnetic fields $\vec{B}$ in the following,

\begin{equation}
\label{L}
L_{aEB}=k_a\alpha \frac{a(\vec{x},t)\vec{E}\cdot\vec{B}}{f_a\pi}
\end{equation}
with the fine structure constant $\alpha\simeq 1/137$,   
where the numerical constant $k_a$ depends on axion models; typically it is of the order of one.
Hereafter we set $k_a=1$. The interaction term slightly modifies Maxwell equations;

\begin{eqnarray}
\vec{\partial}\cdot\vec{E}+\frac{\alpha\vec{\partial}\cdot(a(\vec{x},t)\vec{B})}{f_a\pi}&=0&, \quad 
\vec{\partial}\times \Big(\vec{B}-\frac{\alpha a(\vec{x},t)\vec{E}}{f_a\pi}\Big)-
\partial_t\Big(\vec{E}+\frac{\alpha a(\vec{x},t)\vec{B}}{f_a\pi}\Big)=0,  \nonumber  \\
\vec{\partial}\cdot\vec{B}&=0&, \quad \vec{\partial}\times \vec{E}+\partial_t \vec{B}=0.
\end{eqnarray}
From the equations, we approximately obtain the electric field $\vec{E}$
generated on the axion stars under the background homogeneous magnetic field $\vec{B}$,
\begin{eqnarray}
\label{ele}
\vec{E}_a(r,t)&=&-\alpha \frac{a(\vec{x},t)\vec{B}}{f_a\pi}
=-\alpha \frac{a_0\exp(-r/R_a)\cos(m_at)\vec{B}(\vec{r})}{\pi} \nonumber \\
&\simeq& 1.3\times 10^{-1}\,\mbox{eV}^2( \,\,\simeq0.7\times 10^4\mbox{eV}/\mbox{cm} \,\,)\cos(m_at)\times \nonumber \\
&&\Big(\frac{M_a}{2\times 10^{-12}M_{\odot}}\Big)^2\Big(\frac{m_a}{0.6\times10^{-5}\mbox{eV}}\Big)^3\frac{B}{10^{10}\mbox{G}}\frac{\vec{B}}{B}.
\end{eqnarray} 
with  $r\ll R_a$,
where we have taken into account that the momenta of the axions are vanishingly small; $\vec{\partial}a(\vec{x},t)\simeq 0$.

This is the electric field discussed in the previous section.  It oscillates with the frequency $\nu_{\rm int}=m_a/2\pi$ because
the axion field does with the frequency.
The electric field makes electrons coherently oscillate over the region with the volume $2\pi/m_a$  
in which magnetic field can be regarded as spatially homogeneous one.
Because magnetic fields in accretion disks point in the inside of the disks,
the electric fields point in the same direction so that the coherent dipole radiations are emitted outside the disks.
On the other hand, the magnetic fields of neutron stars point in the outside of the neutron stars but not necessarily point in the direction correctly
perpendicular to their surfaces. Thus, the dipole radiations can be emitted outside the neutron stars.  

\vspace{0.1cm}
Because we obtain the oscillating electric field in eq(\ref{ele}), we estimate 
the critical temperature $T_c$ discussed in the previous section,

\begin{equation}
T_c=\Big(\frac{eE_0}{m_a}\Big)^2\frac{1}{2m_e}\simeq 0.5\times 10^3\mbox{eV}\Big(\frac{eB}{10^{10}\rm G}\Big)^2
\Big(\frac{M_a}{2\times 10^{-12}M_{\odot}}\Big)^4\Big(\frac{m_a}{0.6\times 10^{-5}\rm eV}\Big)^4.
\end{equation}
Coherent radiations emitted from electron gases are terminated at the temperature $T_c$.

\vspace{0.1cm}
\subsection{Ejecta from Accretion Disk}
We should mention that the electrons acquire thermal energies ( e.g. $T_c\sim 10^{3}$eV ) in the axion star collision with neutron stars or accretion disks.
Because such a large amount of the energies is deposited on the surface of the neutron star or accretion disk in a very short time ( $<1$ms ), 
the clumps of the gases absorbing the energies may flow outside the surface. 
More importantly, some ejecta with much larger energies than those of the clumps in the surface can be produced 
inside of the accretion disk.
The energies of the coherent radiations produced in deep inside of the neutron stars or the accretion disks 
are absorbed inside the astrophysical objects themselves. Then,
the gases absorbing the radiations can be ejected outside the astrophysical objects, in particular, geometrically thin accretion disk.
( Only a fraction of the gases absorbing the radiation energies would be ejected from the axion star collisions with neutron stars. )
This is because the energies of the radiations are extremely large as we show below and deposited with a very short time less than $1$ msec.
The ejecta arise repeatedly in the FRB 121102.
Therefore, in the collision between the axion star and the accretion disk, 
the gases with their energies more than the observed radiation energies $\sim 10^{37}$erg may be ejected 
from the accretion disk. 
These ejecta would form the persistent radio source or could power the persistent source observed in the vicinity of the FRB 121102.    
In the next section we estimate energies and durations of FRBs.

\subsection{Tidal Defomation and Orbits of Axion Stars}
\label{3.4}
Axion stars are loosely bounded states of axions. Their binding energies $GM_am_a/R_a\sim 10^{-14}m_a$ are extremely small.
Thus, when they approach neutron stars or accretion disks near black holes, their forms are distorted like long sticks.
In the previous paper\cite{iwazaki4} we have estimated their cross sections and lengths when the axion stars collide the neutron stars.
The cross sections ( lengths ) are given by $\sim 1\rm km^2$ ( $\sim 10^5$km ). Such long sticks directly collide with the neutron stars.
On the other hand, when they collide accretion disks, the situation is different from the case in the neutron stars.  
The axion stars as dark matter form halos in galaxies. Their number densities may grow in the center of the galaxies
such as $r^{\alpha}$ with $\alpha\simeq -1$.
Their motions near the galactic black hole would be Kepler motions whose velocities are determined by the masses of the black holes.
Furthermore, we should note that
the dark matter axion stars may form a disk\cite{darkdisk} which is coplanar with the accretion disk.
The dark matter disk is formed by the gravitational force of the accretion disk. 

As far as their locations $r$ are beyond the distance $r_c$ ( $r>r_c$ ) from the black holes, their forms are spherical without tidal deformation. 
When they are near the black holes ( $r<r_c$ ), their forms are distorted and they form axion streams. The streams occupy the orbits of the 
axion stars. The cross sections of the streams\cite{distance} orbiting the black hole are the same as those of the spherical axion stars.
Here, the critical distance $r_c$ has been estimated\cite{distance} such as $r_c\sim \sqrt{GM/v_rv_a}$ where $M$ denotes the mass of black hole;
$v_r$ does relative velocity between axion star and black hole and $v_a\sim \sqrt{GM_a/R_a}$ average velocities of axions in the axion stars.
In particular, repeating FRB 121102 has been observed to be located in a dwarf galaxy with the mass $\sim 10^8M_{\odot}$. 
We assume the mass $10^4M_{\odot}$ of the black hole in the center of the galaxy. Then, the critical distance is roughly given by $r_c\sim 10^9$km
when $v_r=10^{-3}$.
 
We should note that the axion stars never lose their angular momenta not just like gases in accretion disks. Thus, the axion stars
do not approach the black hole. But just like comets in Oort Cloud, sometimes, some of them located at $r>r_c$ 
would fall toward the black hole and hit the accretion disk. 
Namely, we suppose that some of the axion stars are gravitationally 
kicked by the other axion stars and their orbits suddenly approach the black hole.
Then, the axion stars collide the accretion disk near the black hole, accompanying axion streams.
When they hit the accretion disk, the relative velocities between the axion stars and
the accretion disk may be much smaller than their velocities determined by the Kepler motions. 
Thus, the collision lasts until the axion stream passes away; the axion stream hits the almost same point.
As we show in next section, the radio bursts are emitted by electron gases in the collision point 
and the coherent emissions last only for less than a millisecond.
The emissions are terminated with thermal fluctuations of the electrons. 
  
We would like to make a comment that magnetic fields in the accretion disk at $r>r_c$ may be much smaller 
than ones $\sim 10^{10}$ Gauss we consider in the present paper.  
Their strengths increase as we approach near the black hole in the galaxy.
Thus, the axion stars located at the distance $r>r_c$ far from the black hole may collide the
accretion disks but they hardly lose their energies and angular momenta by radio emissions
because of the weak magnetic fields. It would be difficult to observe the emissions with small energies.

When we consider the axion star collision with the neutron star or accretion disk in the subsequent sections,
we treat the axion streams, i.e. deformed axion stars, as the axion stars.

\section{Burst Energy and Duration}
\label{4}
\subsection{Energy of Coherent Radiation}
We show that the energies of FRBs can reach $10^{40}$erg/s when the axion stars collide neutron stars or accretion disks with strong magnetic fields.
The emission of the FRBs takes place in the region whose volume is equal to $S_a\lambda$; $S_a$ ( $\sim 10^4\rm cm^2$ or larger ) denotes
the surface area of the axion star meeting the magnetized objects and $\lambda$ ( $< 2\pi/m_a$ ) does the skin depth.   
The radiations produced in the region below the depth $\lambda$ cannot be emitted outside of the objects.
Because each electron emits the energy  $\dot{w}\equiv e^2\dot{p}^2/3m_e^2=e^4\vec{E_0}^2/3m_e^2$ 
in the oscillating electric field $\vec{E}=E_0\cos(m_at)$ in eq(\ref{ele}), the electrons in the volume $(2\pi/m_a)^2\times \lambda \sim
(10\rm cm)^2\times 0.1\rm cm$ 
can emit coherent radiations with the energies per unit time,

\begin{equation}
\label{W}
\dot{W} =\dot{w}\bigg(n_e\times 10^2\mbox{cm}^2\times 0.1\mbox{cm}\bigg)^2
\sim 10^{33}\,\mbox{erg/s}\,\Big(\frac{n_e}{10^{22}\mbox{cm}^{-3}}\Big)^2
\,\Big(\frac{M_a}{2\times 10^{-12}M_{\odot}}\Big)^4\Big(\frac{B}{10^{10}\mbox{G}}\Big)^2,
\end{equation}
where $n_e$ denotes electron number density assumed to be large such as $10^{22}\rm cm^{-3}$. 

The coherent radiations are emitted from the region with the volume $(2\pi/m_a)^2\lambda$. 
The surface area of the axion star meeting the magnetized objects is composed of a number of such regions 
$S_a/(2\pi/m_a)^2\sim (10^4\rm cm)^2/(10 \rm cm)^2=10^6$,
where we assume the surface area $(10^4\rm cm)^2$.  
( Probably, the surface area in the actual collisions may be larger than the one. ) 
Therefore, the total radiation energy $\dot{W}_{tot}$ produced in the axion star collision with the neutron star or the magnetized accretion disk
is given by

\begin{eqnarray}
\label{Wt}
\dot{W}_{tot}&=&\frac{S_a}{(2\pi/m_a)^2}\times\dot{w}\bigg(n_e\times 10^2\mbox{cm}^2\times 0.1\rm cm\bigg)^2 \nonumber \\
&\sim& 10^{39}\,\mbox{erg/s}\,\Big(\frac{n_e}{10^{22}\mbox{cm}^{-3}}\Big)^2
\,\Big(\frac{M_a}{2\times 10^{-12}M_{\odot}}\Big)^4\Big(\frac{B}{10^{10}\mbox{G}}\Big)^2.
\end{eqnarray} 
The skin depth $\lambda$ is approximately given  
such that $\lambda\sim \sqrt{1/\nu_{\rm int}\sigma}$ with the electric conductivity of electrons
$\sigma=\omega_p^2\tau/4\pi$, where $\omega_p=\sqrt{e^2n_e/m_e}$ denotes
the plasma frequency and $\tau$ does the mean free time of the electrons. Numerically, they are given by $\lambda\sim 0.1$cm,
$\omega_p\sim 10^{16}$s$^{-1}$ and $\tau\sim 10^{-15}$s for $n_e=10^{22}$/cm$^3$ and temperature $\sim 10^6$K. 
( The detail of the mean free time $\tau$ is discussed soon below. )
The radiations produced in the region deeper than the skin depth are absorbed in dense electron gases\cite{hamada}.
We find that the emission mechanism of the FRBs can explain the observed large amounts of their energies  $\sim 10^{40}$erg/s.

The surface area $S_a$ of the axion stars meeting the neutron star or magnetized accretion disk
is assumed to be $(10^4\rm cm)^2$, although
they can be larger than $(10^4\rm cm)^2$ even if
the axion stars are distorted\cite{iwazaki4} by tidal forces of neutron stars or black holes. 
( We have shown\cite{iwazaki4} that the axion stars become to look like long sticks owing to the tidal forces of neutron stars. ) 
In particular, when the axion stars collide the accretion disks, their deformation by the tidal forces of the black holes 
would be not so large compared with the case in the direct collisions with the neutron stars. Thus, the surface area is much larger than the value
taken in the above estimation. 

In general, it is well known that the magnetic fields of the neutron stars are stronger than $10^{10}$G. Thus, the radiation energies 
are much larger than the value in eq(\ref{Wt}). On the other hand, it is not clear how strong are
the magnetic fields in the accretion disks of the black holes. So tentatively we use the value $10^{10}$G in order to
explain the radiation energies of the repeating FRB 121102.


\vspace{0.1cm}
The number density $n_e$ of electrons used in the estimation should be regarded as
the average one. The number density of the electrons exponentially increases
from the values ( $\simeq 0\,$cm$^{-3}$ ) in vacuum to the values ( $>10^{22}$cm$^{-3}$ ) in the surfaces of the objects. In particular, 
the number density in atmospheres with low temperatures of neutron stars increases 
from $\sim 0\,$cm$^{-3}$ in the vacuum to the one much larger than $10^{22}$cm$^{-3}$ 
even at the skin depth $0.1$cm. Therefore, the value taken in the above estimation should be regarded as the average one.
As we discussed above, the radiations from the surfaces of the neutron stars can be emitted outside of them.
The results have been confirmed more rigorously in our previous paper\cite{iwazaki5}, in which
we have calculated the opacity of the electron gases in the atmospheres of neutron stars.

The radiations emitted from the geometrically thin accretion disks are expected to be not absorbed after their emissions.
Their emissions arise at the surfaces of the accretion disks near the galactic black holes where the disks are very thin. 
The number densities of electrons also exponentially decrease toward the outside 
when the gas pressure balances with the gravitational pressure in the disks. 
However, because the structures of the geometrically thin accretion disks
are less well understand than the one of the neutron star atmospheres, 
we simply assume that the radiations emitted from the disks are not absorbed.

In any case, we find that there are physically allowed parameters with which the large radiation energies ( fluence ) can be explained
in the axion star model for the repeating and non repeating FRBs.   

\vspace{0.1cm}
A few comments are in order.

First,
an emission mechanism in the collisions between axion stars and neutron stars has been discussed in the references
\cite{conversion,hamada}. The mechanism is the conversion
of the axions into radiations under strong magnetic fields of the neutron stars. 
One axion is converted to one photon.
The radiations are incoherent. Thus,
the energy fluxes are not sufficiently large to be detectable 
when the collisions occur in extragalactic universe. 
Thus, the fluxes are proportional to the number of the axions in the axion star. Although the fluxes are not large,
the radiations show the line spectrum with the frequency $m_a/2\pi$ so that we can directly determine the mass of the axions.

\vspace{0.1cm}
Secondly,
in the estimation of the amount of the radiation energies, we have used the mixing $\propto \alpha\, a(\vec{x},t)\vec{E}\cdot\vec{B}/f_a$ in eq(\ref{L})
between the axion and the photon. Especially, the coefficient of the term derived in the vacuum has been exploited. But the coefficient
has been shown\cite{suppresion} to be suppressed in the dense plasma by the factor $(m_a/\omega_p)^2$. The suppression is caused by the Ohm's law.
That is, the Ohmic dissipation of the energy of the electric field $\vec{E}$ effectively diminishes the mixing in the dense plasma.
But, our mechanism of the coherent dipole radiations works before the beginning of the suppression due to the Ohm's law.
The radiations are emitted by the coherently oscillating electrons and their emissions last until the thermal fluctuations disturb the oscillations.
Such large thermal fluctuations are realized by the thermalization of the oscillating energies. 
That is, the currents of the oscillating electrons are suppressed due to the Ohmic dissipation after
the electrons interact with each other many times. 
In other words, the coherent radiations last 
for the mean free time of the electrons. Therefore, our use of the mixing in the vacuum is valid for the above estimation.

\vspace{0.3cm}
\subsection{Duration of Coherent Radiation}
The emissions of the FRBs are stopped by thermal fluctuations. Namely, the oscillation energies are thermalized so that the temperatures of the electron gases increase.
Eventually, the thermal fluctuations disturb the coherent radiations. The critical temperature $T_c$ at which the 
coherent emissions stop is determined by the oscillation energies $p^2/2m_e\simeq (eE_0/m_a)^2/2m_e$;
the thermal energies at the temperature $T=T_c$ are equal to the oscillation energies.
Then, the durations of the bursts may be approximately given by the mean free time of electrons at the temperature $T_c$. 
So we roughly estimate the mean free time $\tau$ of electrons. 
Electrons interact with each other by the Coulomb interaction. When the cross section of the electrons is $\pi l^2$, the mean free time $\tau$ is given by
$\tau=1/(\pi l^2 n_e v)$ where the velocity $v$ of electrons is $\sqrt{2T_c/m_e}$. The cross section $\pi l^2$ is estimated such that 
the distance $l$ between electrons satisfies
$\alpha/l\simeq T_c\simeq (eE_0/m_a)^2/2m_e$; Coulomb energy is equal to kinetic energy.
Thus, the mean free time is given by

\begin{equation}
\label{tau}
\tau(T_c)\simeq \frac{\sqrt{m_e}T_c^{3/2}}{\sqrt{2}\pi\alpha^ 2n_e}\sim 10^{-9}\mbox{s}
 \Big(\frac{T_c}{10^5\mbox{eV}}\Big)^{3/2}\Big(\frac{10^{22}\mbox{cm}^{-3}}{n_e}\Big),
\end{equation}
where we tentatively used the number density $n_e\sim 10^{22}/\rm cm^{3}$ of electrons and the critical temperature $T_c=10^5$eV.
( The critical temperature $T_c$ depends on the magnetic field $B$; $T_c\sim 10^5\mbox{eV}(eB/10^{11}\rm G)^2$.)
We may consider that the mean free time of electrons is roughly equal to the intrinsic duration of the coherent radiations. 
This intrinsic duration is much less than the observed values. Owing to the propagation
effects by electron gases in intergalactic space or near the source of the FRBs, the observed durations ( pulse widths ) become much longer than 
the intrinsic one. In this way, we find that
the observed durations $\sim 1$ms of the FRBs are consistent with the above estimation in the axion model.
( We have not taken into account the thermal effects on the number densities $n_e$ in the above calculations. In general, the number densities $n_e$ depend
on the temperatures. Thus, the densities decrease as the temperatures increases. Then, the intrinsic durations $\tau$ become longer. )

\section{Spectral and Temporal Properties of FRBs}
\label{5}
\subsection{Center Frequencies and Bandwidths of Repeating FRBs} 
Now, we analyze the properties of FRBs, especially, the detailed properties of the FRB121102 by using the axion model.
First, the spectra of the repeating FRB 121102 have been recognized to be narrowband, not broadband.
The spectra are characterized with two parameters; center frequency $\nu_c$ and bandwidth $\delta\nu$.
According to the axion model, the line emissions are broadened by thermal effects. Such spectra are given by

\begin{equation}
\label{S}
S(\nu)\propto \exp\Big(-\frac{(\nu-\nu_c)^2}{2(\delta\nu)^2}\Big),
\end{equation}
with the bandwidths $\delta\nu=\nu_c\sqrt{T_c/m_e}$. We notice that the bandwidths are proportional to the center frequencies $\nu_c$.
This is a property resulted from the effects of the thermal fluctuations on the line spectra. 

Such a proportionality has been observed to be roughly valid,

\begin{eqnarray}
\label{10}
\delta\nu&\sim& 1000\mbox{MHz} \quad \mbox{for center frequencies}\quad \nu_c\sim 6\mbox{GHz in the ref.\cite{4-8}} \nonumber \\
\delta\nu&\sim& 500\mbox{MHz} \quad \mbox{for center frequencies}\quad \nu_c\sim 3\mbox{GHz in the ref.\cite{repmulti1}} \nonumber \\
\delta\nu&\sim& 300\mbox{MHz} \quad \mbox{for center frequencies}\quad \nu_c\sim 2\mbox{GHz in the ref.\cite{repmulti}} \nonumber \\
\delta\nu&\sim& 200\mbox{MHz} \quad \mbox{for center frequencies}\quad \nu_c\sim 1.2\mbox{GHz in the ref.\cite{rep1}}.
\end{eqnarray}

Although the observations exhibit that each width $\delta\nu$ with an almost identical center frequency vary depending on each burst, the above values have been taken as typical ones. 
( Indeed, the variations in the widths are not large; for instance, the variations is given such that $400\rm MHz<\delta\nu<700MHz$ for the center frequency $\nu_c\simeq 3$GHz
\cite{repmulti1} whose bandwidth is referred as $\sim 500$MHz. )
The proportionality is never strict, but we can see such a tendency in the spectra.
The theoretical prediction by the axion model is that 
the proportional constant $\sqrt{T_c/m_e}=eE_0/(\sqrt{2}m_am_e)\propto (M_am_a)^2B$ depends on the strength of the magnetic field $B$.
( It also depends on the masses $M_a$ of the axion stars, which may take a value in a range $(10^{-12}M_{\odot}\sim 10^{-11}M_{\odot})$. )
Thus, the variation in the widths $\delta\nu$ at an identical center frequency comes from the variation in the magnetic fields or the masses of the axion stars.

It is interesting that
the widths $\delta\nu$ are estimated such as ,
 
\begin{equation}
\label{11}
\delta\nu=\nu_c\sqrt{\frac{T_c}{m_e}}=\frac{\nu_c eE_0}{\sqrt{2}m_am_e}=590\mbox{MHz}\,\frac{\nu_c}{3\mbox{GHz}}\,\frac{eB}{5\times10^{10}\rm G}
\Big(\frac{M_a}{2\times 10^{-12}M_{\odot}}\Big)^2.
\end{equation} 
where we take the strength of 
the magnetic fields $eB\simeq 5\times 10^{10}$G.
It is remarkable that by using the almost identical strengths of the magnetic fields $B= (10^{10}\rm G\sim 10^{11}G)$ for the evaluation of
the burst energies $\dot{W}_{tot}$, durations $\tau(T_c)$ and bandwidths $\delta\nu$, we can obtain 
corresponding observed values of the FRBs.

\vspace{0.2cm}
Now, we proceed to answer why there are various center frequencies $\nu_c$; $1\rm GHz<\nu_c<  7$GHz. It apparently seems that
the frequency of the dipole radiations is uniquely given by $\nu_{\rm int}=m_a/2\pi$ in the axion model.
We should note that the radiations of the FRB 121102 are emitted from the electron gases in the accretion disks 
which the axion stars collide. The disks can closely orbit a galactic black hole
and then their velocities $\vec{V}$ can be relativistic. Thus, the radiations emitted from the disks are Doppler shifted. The Doppler shift is given by 
$\nu_c(V)=\nu_{\rm int}\Pi(V)$ with $\Pi\equiv\sqrt{1-V^2}/(1-V\cos\theta)$ where $\theta$ denotes the angle between the direction of $\vec{V}$ and the line of sight.
Therefore, the frequencies $\nu_c$ can take large values such as $6$GHz when $V=|\vec{V}|\simeq 0.95$ and $\theta \simeq 0$
when $\nu_{\rm int}=m_a/2\pi \simeq 1.4$GHz$\big(m_a/0.6\times 10^{-5}\rm eV\big)$. Various velocities $\vec{V}$ of the disks give rise to
the various center frequencies $\nu_c$.
This is the reason why there are bursts with various center frequencies in the repeating FRB 121102.
We can predict from the discussion that the bursts with frequencies much larger than $10$GHz are absent because
the velocity of the disks from which such radiations are emitted, must be almost equal to light velocity. The axion star collisions 
with such disks might be extremely rare.

The recently observed repeating FRB 180814.J0422+75 shows the lower center frequencies $\sim 700$ MHz than the ones the FRB 121102 shows. 
This implies that the axion star collisions takes place with the angle $\theta\sim \pi$
in the FRB 180814.J0422+75.

\vspace{0.1cm}
The Doppler shifts also give rise to a temporal property pointed out in the reference \cite{4-8} that durations $\tau(\nu_c)$ of bursts with higher center frequencies $\nu_c$
are shorter than those with lower center frequencies.
The durations shorten owing to the Doppler effect, that is, $\tau=\tau_{\rm int}\Pi^{-1}(V)$. The duration $\tau_{\rm int}$ denotes 
the duration measured by observers moving with the disks.
So, $\tau(\nu_c)<\tau(\nu'_c)$ when $\nu_c(V)>\nu'_c(V')$, that is $V>V'$.
It is obvious that the product of $\nu_c$ and duration $\tau(\nu_c)$ does not receive the Doppler effect.
The feature can be seen in the Fig.7 of the reference\cite{4-8}. Our argument holds under the assumption that the durations observed at the earth are 
proportional to the durations observed near the source of the radiations: 
The propagation effects on the durations do not change the result.
( Actually, it has been discussed\cite{pulse} that the broadened pulse widths caused by intergalactic medium effects are 
smaller than the intrinsic ones discussed in the paper. This is because
the location of the galaxy involving the source of the FRB 121102 is near the Milky Way Galaxy, that is, its redshift is $z\simeq 0.2$. ) 

\vspace{0.1cm}
( It appears that the Doppler effect gives rise to larger peak flux densities $S(\nu_c)$ 
as the center frequencies become higher; $S(\nu_c)=(\nu_c/\nu'_c)^3S(\nu'_c)$. The Doppler effect make flux densities $S_0$ large
such as $S=S_0\Pi^3$. The effect on $S$ is much larger than the effect on $\tau$ or $\nu_c$.
But the trend cannot be
observed\cite{4-8} in the FRB 121102. As pointed out\cite{4-8}, the distribution of the peak flux densities are almost flat in the range $1\rm GHz \sim 6GHz$.
Probably, the flat distribution comes from the fact that both of 
the electron number densities $n_e$ and the strength of magnetic fields $B$ become smaller\cite{accretion,review} 
as the accretion disks closer approach the black hole, for instance,
$n_e\propto r$ and $B\propto r$ where
$r$ denotes the radial coordinate measured from the center of the black hole.
Then,
because $\dot{W}$ are proportional to $(Bn_e)^2\propto r^4$, the observed flux densities $S(r)\propto \dot{W}(r)$ emitted from the disks at $r$ 
can be expressed by 
those $S(r')$ emitted from disks at $r'$ ( $r'>r$ ) such that $S(r)=(r/r')^4(\nu_c/\nu'_c)^3S(r')$. Note that $\nu_c(r)>\nu'_c(r')$. 
The factor $(r/r')^4$ would diminish the influence of the Doppler effect, i.e. $(\nu_c/\nu'_c)^3$ on the peak flux densities. 
Thus, the distribution of the
observed flux densities must be almost flat. We should make a comment that 
the dependences of $V(r)$ and $n_e(r)$ on $r$ depend on each model of the accretion disks. So we can
not precisely estimate the factor $(r/r')^4$.  )

\vspace{0.2cm}
\subsection{Difference between Repeating FRBs and Non-repeating FRBs}
Non repeating FRBs are generated by the axion star collision with neutron stars, while repeating FRBs are generated by its collision with accretion disks of
black holes. Both objects have strong magnetic fields. In general, we expect that the magnetic fields of the neutron stars are stronger than those of the accretion disks.
Stronger magnetic fields produce larger flux densities of FRBs because $\dot{W}$ in eq(\ref{W}) is proportional to $B^2$. 
Actually, the flux densities of the non repeating FRBs are
roughly ten times larger than those of the repeating FRB 121102. The difference may come from the difference in the strengths of
the magnetic fields 
in the astrophysical objects. Obviously, the flux densities also depend on the electron number densities $n_e$.
( $\dot{W}$ is proportional to $n_e^2$. ) The average number densities of electrons producing FRBs in the neutron stars 
might be larger than corresponding number densities of electrons in the accretion disks.
Thus, we cannot determine which one is the main cause leading to the difference in the flux density.
Probably, both causes make a difference in flux densities of non repeating FRBs and the repeating FRB.

\vspace{0.1cm}
Furthermore, the difference in the magnetic fields gives rise to the difference in the bandwidths of FRBs.
The bandwidths depend only on the magnetic fields but not on electron number density $n_e$; $\delta\nu$ in eq(\ref{11}) is proportional to $B$. 
We expect that the bandwidths of the non repeating FRBs are much larger than those of the repeating FRB.
Indeed, the bandwidths of non repeating FRBs have not yet been detected. They would be too large to be in the extent of receiver frequency ranges.
They have been observed only in the frequency range $0.8\rm GHz\sim 1.6GHz$. ( People might think that the bursts are broadband. )
The bandwidths of the non repeating FRBs would be $3\sim 5$ times larger than those of the repeating FRBs. Namely, 
$\delta\nu$ of the non repeating FRBs may be $600\rm MHz\sim 1$GHz at the center frequency $1.4$GHz. 
The bandwidths are over the extent of the receiver frequency range.
 ( The very recent MWA observations\cite{finite} at the frequency $200$MHz have not detected non repeating FRBs, although they have been detected
by simultaneous ASKAP observation at frequency $1.4$GHz. It suggests that the non repeating FRBs are not broadband and their bandwidths are less than $1.2$GHz. )
If observations with the wide extent of the receiver frequency range 
( e.g. a frequency range $1\rm GHz \sim 3$GHz )
are performed, the bandwidths of the non repeating FRBs could be observed.

%

\vspace{0.2cm}
\subsection{Non Periodicity of Repeating FRBs}
The repeating FRB 121102 takes place
when the axion stars collide the accretion disk of a galactic black hole.
The host galaxy has been identified; the mass is estimated to be $\sim 10^8M_{\odot}$.
It is reasonable to assume that the mass of a black hole in the galaxy is about $10^4M_{\odot}$.
Many axion stars as dark matter may be present at the center of the galaxy. 
Especially, their motions may be described by the Kepler motions so that the
velocities are determined by the mass of the nearby black hole. 
As we mentioned in the previous section (\ref{3.4}), the axion star dark matter may form a disk coplanar to the accretion disk.
When the axion starss are located at the distance larger than the critical distance $r_c\sim 10^9$km, the effects of the tidal force of the black hole are ineffecient.
But, once they approach the black hole ( $r<r_c$ ), their forms are destroted so that they behave as axion streams.
The cross sections of the streams are the same as those of the axion stars. 
The streams may occupy their orbits. 
On the other hand, when an axion star located at $r>r_c$ is gravitationally kicked by the other axion stars, it falls to the orbit near the black hole
and collides the accretion disk.
Thus, the collision with the disk may irregularly occur just like the emissions of the repeating FRB 121102.
After it collides the disk, it would orbit the black hole and then collides again. Finally it disappears with evaporation.
Then, there would be several collisions. 
If the velocity of the axion star is of the order of $\sim 10^5$km/s with
the radius of the orbit $\sim 10^5$km,
the sequence of the collisions is of the order of $1$ second.
( In the case of the larger radius than $10^5$km, the velocity is lower than $10^5$km/s and the sequence is much longer than $1$ second. )
These are consistent with the observation\cite{periodicity} of the FRB 121102.
After such collisions, the axion star loses its energy and evaporates.
Subsequent collisions would occur one hour or one month later when the next axion star is kicked to fall toward the black hole.
It is unpredictable.
 
Some of the bursts in the FRB 121102 exhibit a few peaks in
their spectra\cite{4-8}. That is, $2$ or $3$ peaks appear within a few msec. 
They might be caused by the splitting of the axion star. The axion stream has a cross section $\sim (10^2\rm km)^2$.
A part of the stream collides the accretion disk first and then a different part collide the different position in the accretion disk.
The time difference is a few msec because the velocity is of the order of $10^5$km/s.
Then, $2$ or $3$ peaks appear in the spectrum. 
 
Although the complexity in the time sequence is present, 
we can see that the distribution\cite{periodicity} of the time intervals between sequent bursts
is concentrated in shorter span and nonvanishing even in longer span. 
The concentration might be caused by the axion stars falling into the black hole as they orbit, as we have shown. 

The observations\cite{periodicity} also exhibit many bursts with low flux densities $\sim 50$mJy of compared with
the flux densities previously observed\cite{4-8}. It suggests that the bursts with low flux densities arise from the axion stars with 
smaller masses than the one ( $\sim 2\times 10^{-12}M_{\odot}$ ) referred in the present paper. 
Probably, the axion stars with smaller masses might be produced by the splitting of original axion stars with masses $\sim 2\times 10^{-12}M_{\odot}$.

\vspace{0.1cm}
\subsection{A New Repeating FRB 180814.J0422+75}
It has recently been reported\cite{chime} that a new FRB 180814.J0422+75 
with low frequencies $400\rm MHz\sim 800$MHz has been observed. It is interesting that the spectra of all repeating bursts 
with such low frequencies are narrowband. Furthermore,
the bandwidths of the bursts with the center frequencies $\sim 700$MHz  ( $450$MHz )
are approximately given by $90$MHz ( $50$MHz ). The result is coincident with our results mentioned above;
the bandwidths are proportional to the center frequencies. More interestingly, we can see the feature that pulse widths ( durations ) 
are broader as the center frequencies are lower. The feature is identical to the one observed in FRB 121102.
All of these properties of the FRB 180814.J0422+75 are caused by Doppler effects on the radiations emitted by the accretion disks
of galactic black holes. 

CHIME\cite{chime2} has observed $13$ bursts, one of which is the repeating FRB 180814.J0422+75. According to our axion model,
the bursts with the finite widths and their maximum frequencies much less than $800$MHz are
necessarily repeating FRBs. At least, the bursts are emitted by the accretion disks. Similarly, the bursts with the lowest frequencies
much higher than $2$GHz, the bursts are repeating FRBs.

The repeating FRBs arises from the collisions between the accretion disks and the axion stars. 
The angle $\theta$ between the velocity $\vec{V}$ of the accretion disk hit by the axion star
and the line of sight can take values $0<\theta<\pi$. 
When $\theta\sim 0$ ( $\theta\sim \pi$ ), the burst frequencies are higher ( lower ) than $2$GHz ( $800$MHz ).
Therefore, we predict that the bursts with low frequencies such as $600$MHz are
observed for FRB 121102, while the bursts with high frequencies such as $3$GHz are observed for the FRB 180814.J0422+75.
( The prediction has been confirmed in the recent observation\cite{lowre}. )

\section{Discussion and Summary}
\label{6}

As we have mentioned,
the rotation measures of some FRBs have been obtained\cite{rotation,lrotation}. Among them, 
the rotation measures\cite{lrotation} of the FRB 121102 are very large such as $\sim 10^5$rad m$^{-2}$, while
those of the non repeating FRBs are at most $\sim 2\times10^2$rad m$^{-2}$. It indicates that the environment
of the FRB121102 is similar to the one of PRS J1745-2900 located near the black hole in 
the Milky Way galaxy, 
while those of the non repeating FRBs are similar to environments of ordinary pulsars in the Milky Way galaxy.
We should note that recently observed two non repeating FRB190523\cite{523} and FRB180924\cite{924} 
are localized to ordinal galaxies similar to the Milky Way. It
is completely consistent with our axion model; non repeating FRBs arise from neutron stars, while the repeating FRBs do from
accretion disks of black holes in dwarf galaxies.

\vspace{0.1cm}
We have analyzed spectral and temporal properties of FRBs
by using the axion model. The model states that the axion star collisions with neutron stars cause non repeating FRBs, while their collisions with 
accretion disks of galactic black holes cause repeating FRBs. Electron gases on the surfaces of the astrophysical objects emit coherent radiations 
when the axion stars collide them. This is because the axion stars generate spatially homogeneous oscillating electric fields
when they are under homogeneous strong magnetic fields. The electric fields make electrons coherently oscillate and emit the coherent radiations.
The homogeneity should hold over the region with the volume $S_a\lambda$ such as $S_a\lambda\sim (10\rm cm)^2\times 0.1\mbox{cm}\simeq 10\rm cm^3$,
from which the coherent radiations are emitted.
But, the coherent radiations stop because the thermal fluctuations owing to the thermalization of the oscillation energies
disturb the coherent oscillations of the electrons. We have shown that the durations of the FRBs
are much shorter than $1$ms. The thermal effects make the spectra of the radiations broaden.

Because the accretion disks can orbit with relativistic velocities in the vicinity of the black holes, the radiations emitted from the disks are Doppler shifted.
Accordingly, the radiations with higher center frequencies ( $2\rm GHz\sim 8$GHz )
than the intrinsic frequency $m_a/2\pi\simeq 1.4$GHz$(m_a/0.6\times 10^{-5})$eV are observed in the repeating FRB 121102. 
Similarly, these bursts with higher frequencies exhibit shorter durations than those of bursts with lower frequencies due to the Doppler effects.
On the other hand, the non repeating FRBs arise from the collisions with neutron stars.
There are no Doppler effects which produce various center frequencies. Thus, there are no 
non-repeating FRBs with higher frequnecies $>3GHz$ or lower frequencies $<500$MHz. 

Non repeating FRBs show larger flux densities than those of the repeating FRB 121102. It may be caused by stronger magnetic fields
of neutron stars than the ones of the accretion disks. Such strong magnetic fields
produce large bandwidths of non repeating FRBs so that they are over the extent of the present receiver frequency range. 
This is the reason why the spectra of the repeating FRBs have been found to be narrowband, 
but narrowband non repeating FRBs have not yet been observed.

We have discussed that in addition to the coherent radiations emitted from the surfaces of accretion disks, 
ejecta with the energies such as $10^{39}$erg are produced in the axion star collisions with the accretion disks.
They arise owing to the absorption of the coherent radiations by the disks themselves; the radiations are 
produced inside of the disks.
In particular, these ejecta repeatedly take place in the repeating FRB 121102. The ejecta would form the observed persistent radio source or could be
an energy source of the persistent radio source associated with the FRB 121102.

The axion stars are coherent states of gravitationally loosely bounded axions and their forms are easily
deformed by tidal forces of neutron stars or black holes. Then, they are not spherical but look like axion streams. 
When they collide neutron stars, the cross sections of the streams is of the order of $1\rm km^2$, while 
they are almost same as those of the spherical axion stars when they orbit the black hole.
The fact may cause
the complex structures such as $2$ or $3$ peaks in the spectra within a few msec. 

After finishing the paper, the new repeating FRB 180814.J0422+75 was found. 
The new repeating bursts have lower center frequencies than $1$GHz contrary to those of the FRB 121102.
It is remarkable that their spectral properties are coincident with our results mentioned above; 
narrow bandwidths $\delta\nu$ proportional to center frequencies $\nu_c$, their proportional constant $\delta\nu/\nu_c\sim (0.1\sim 0.2)$ roughly 
coincident with the one of the FRB 121102 and
pulse widths $\tau$ proportional to $\nu_c^{-1}$.

We would like to add a new prediction of the axion mass which can be derived by the recent observation\cite{H} of the FRB121102.
The observation shows the presence of the bursts with the center frequency $\simeq 1.4$GHz with narrow bandwidths less than $100$MHz
and with the low fluence $\sim 0.03$Jy ms.
The bursts are thought to be emitted in the accretion disks with relatively low velocities. This is because the narrow bandwidths 
and the low fluences imply that 
they are produced in the gases with relatively weak
magnetic fields. Those magnetic fields are located in the accretion disks relatively far away from the galactic black hole.  
Because the source of the bursts is located at $z\simeq 0.193$, the intrinsic frequency $m_a/2\pi$ of the bursts is
nearly equal to $1.67$GHz; the axion mass $\simeq 7.16\times 10^{-6}$eV.
 
\vspace{0.1cm} 
According to the results obtained in the paper, 
we conclude that the axion model for the FRBs well explains most of all features of non repeating and repeating FRBs.

\section*{Acknowledgment}
The author
expresses thanks to Prof. Weltman for useful comments, and members in the theory group at KEK for useful comments
and discussions.
This work was supported in part by Grant-in-Aid for Scientific Research ( KAKENHI ), No.19K03832.

\end{document}